\documentstyle[12pt,epsf]{article}

\newcommand{\be}{\begin{equation}}
\newcommand{\ee}{\end{equation}}
\newcommand{\vs}{\vspace{3mm}}
\newcommand{\hs}{\hspace{8mm}}

\begin{document}
\begin{center}

{\large {\bf Quantum Cooperation of Two Insects}}

Johann Summhammer\footnote{Electronic address: 
summhammer@ati.ac.at}

Vienna University of Technology\\
Atominstitut, Stadionallee 2, A--1020 Wien, Austria

\end{center}

{\small {\bf Abstract:}
The physical concept of quantum entanglement is brought to the biological domain. 
We simulate the cooperation of two insects by hypothesizing that they share a 
large number of quantum entangled spin-$\frac{1}{2}$ particles. Each of them makes
measurements on these particles to decide whether to execute certain actions. In 
the first example, two ants must push a pebble, which may be too heavy for one 
ant. In the second example, two distant butterflies must find each other. In 
both examples the individuals make odour-guided random choices of possible 
directions, followed by a quantum decision whether to push/fly or to wait. With 
quantum entanglement the two ants can push the pebble up to twice as far as 
independent ants, and the two butterflies may need as little as half of the 
flight path of independent butterflies to find each other.
}

\vs

{\bf Contents}

1. Introduction

2. Quantum entanglement (QE)

\hs 2.1. Physical background

\hs 2.2. Possibility of QE in the biological domain

3. The models

\hs 3.1. Two ants pushing a pebble

\hs 3.2. Two butterflies finding each other

4. Discussion

\hs 4.1. Communication

\hs 4.2. Strength of correlations

\hs 4.3. Critique of the models

\hs 4.4. Creation and persistence of QE

5. Conclusion

\hs Acknowledgment

\hs References

\vspace{5mm}

{\bf 1. Introduction}

A good part of the communication between the members of a species serves to coordinate their behavior
in the interest of common survival. It is generally believed that this communication is governed by
the laws of classical physics. Examples would be sound, vibration and direct touch,
molecular signalling in the form of smell, and the wide field of behavioral expression, which is physically
a method of modulating or emitting patterns of electromagnetic radiation. However, in the newly emerging branch of 
physics called quantum information \cite{Nielsen} it has become clear that many tasks requiring coordination 
between the actors can be achieved significantly better if the actors' decisions are quantum entangled. The basis 
for this is Bell's theorem, which proves that observational results obtained at two widely separated but quantum 
entangled sites can exhibit correlations whose magnitude surpasses that of any correlations conceivable by 
classical physical laws \cite{Bells_Theorem}.
Given the importance of correlated action between living systems it is worth while to investigate
how quantum entanglement could be embedded beneficially in the stream of sensing, deciding and acting 
of individuals. In this paper we do this by means of two examples. We show how much farther two cooperating ants 
could push a heavy pebble, and how much faster two distant butterflies could find each other.
Since the quantum entanglement is vulnerable in a thermal environment, the models incorporate the quantum 
entanglement in the behavior of the individuals in a way which enables them to solve the cooperative task even 
if the entanglement breaks down, although with less efficiency.

\vspace{5mm}

{\bf 2. Quantum Entanglement (QE)}

\vs

{\em 2.1. Physical background}

In physics, quantum entanglement (QE) is a statistical correlation 
between the properties measured on two or more separated particles or systems. The theory knows no upper limit on 
the complexity and on the number of the systems for which QE can  exist. If only two particles or systems are 
involved, QE means that the result of a measurement done
on one of them is not independent of the result of the measurement done on the other one, no matter how far 
they are apart. While correlations as such are not surprising, the particular aspect of QE correlations is their 
sheer strength, which can give the impression that each particle {\em 'knows'} which measurement result has 
been obtained, or will be obtained, on the other one, so that it can {\em 'behave'} accordingly in its own 
measurement. 
Einstein called these correlations {\em 'spooky action at a distance'} \cite{EPR_original}. Meanwhile, numerous
experiments have shown the existence of these correlations, including tests which demonstrated that they are 
faster than light \cite{Zeilinger_Innsbruck_EPR}. Mathematically, these correlations are outside of time and are 
thus instantaneous. Nevertheless, it is not possible to use QE for the transmission of information, because the 
individual measurement results are unpredictable. A popular but concise 
account of the essence of QE can be found in \cite{Mermin_1985}.

In this paper we make use of the simplest kind of QE, which is the correlation of the angular momentum between 
two particles of the same kind. Both shall have an angular momentum of $\hbar/2$, and are thus so called 
spin-$\frac{1}{2}$ particles. $\hbar$ is Planck's constant of action.
The quantum measurement of the angular momentum is somewhat different from the measurement of angular
momentum of a macroscopic object like a spinning wheel. The quantum measurement consists in choosing a spatial 
direction and determining the value of the angular momentum of the particle along this direction. 
A spin-$\frac{1}{2}$ particle can only yield
two different results, independent of the chosen direction: either $+\hbar/2$ or $-\hbar/2$. For simplicity, these
results are called {\em spin up} ($'+'$) and {\em spin down} ($'-'$), respectively. The angular momenta of two 
such particles can easily be measured along different directions, when the particles are sufficiently far apart. 
The possible results are then $'++'$, $'--'$, $'+-'$, and $'-+'$. Quantum theory can only predict the 
probabilities for these results, $p_{++}$, $p_{--}$, etc. What these probabilities look like depends on the 
particular physical state. An important state, which will be used in the example with the butterflies 
below, is the so called {\em singlet} state. Here, the angular momenta of the two particles are always opposite
to each other, so that both of them together have angular momentum zero. Symbolically, this state is written as
\be
| \psi\rangle_s = | +- \rangle - | -+ \rangle.
\ee
The probabilities for the possible measurement results are 
\be
p^{(s)}_{++}=p^{(s)}_{--}=\frac{1}{2}\sin^2\left(\frac{\alpha}{2}\right),
\ee
\be
p^{(s)}_{+-}=p^{(s)}_{-+}=\frac{1}{2}\cos^2\left(\frac{\alpha}{2}\right),
\ee
where $\alpha$ is the angle between the two chosen directions of measurement. Note that these expressions do not
depend on the {\em individual} directions along which the angular momenta are measured. The time of the measurements
is also not important. The particles need not be measured at the same time, and it is irrelevant which is measured 
first. Moreover, the distance of the particles plays no role. That is why these correlations are called 
{\em nonlocal}.

If one looks at the measurement results of only one particle, e.g. the first one, the probability to obtain
$'+'$ is simply the probability to obtain $'+'$ on the first and $'+'$ on the second particle, plus the probability
to obtain $'+'$ on the first and $'-'$ on the second one. Hence, 
$P_+^{(1)} \equiv p_{++}+p_{+-}=1/2$. Similarly for $'-'$: $P_-^{(1)} \equiv p_{-+}+p_{--}=1/2$. The same holds, if
one looks only at the second particle. The result of a measurement on a single particle is thus completely random, independent
along which direction the angular momentum is measured. In our biological models this will mean that the behavior of each 
individual will be random, except for the guidance given by smells, but the random moves of both individuals will be quite coordinated.
 
Another important state of QE of two spin-$\frac{1}{2}$ particles is the so called triplet state,
\be
| \psi\rangle_t = | ++ \rangle + | -- \rangle,
\ee
which will be used in the examples with the ants below.
Here, the angular momenta of the two particles point always in the same direction, giving a total angular momentum of $\hbar$.
The probabilities for the possible measurement results are:
\be
p^{(t)}_{++}=p^{(t)}_{--}=\frac{1}{2}\cos^2\left(\frac{\alpha}{2}\right),
\ee
\be
p^{(t)}_{+-}=p^{(t)}_{-+}=\frac{1}{2}\sin^2\left(\frac{\alpha}{2}\right).
\ee

\vs

{\em 2.2. Possibility of QE in the biological domain}

As a by-product of interactions, QE should be an omnipresent feature in nature. However, this would mostly be
at the molecular level and below. In systems of macroscopic scale, be they inanimate or alive, one would
expect that the effects of QE would quickly be lost, because any macroscopic system interacts permanently with 
the environment through the exchange of thermal radiation. And as soon as such interaction with systems
outside the original QE occurs, the original QE gets diminished or fully destroyed.
This is partly countered by
the fact that, with increasing complexity of the entanglement --- as would be the case in systems of many degrees 
of freedom, e.g. from molecules upwards --- the deviation from classical 
physics as witnessed by violations of Bell inequalities becomes stronger 
\cite{Marek},\cite{Mermin}. Therefore, traces of QE might be noticeable between such systems even after appreciable
contact with the environment. In certain solids this has already been detected, because macroscopic properties
like the behavior of the magnetic susceptibility are a proof of entanglement even at finite 
temperature \cite{caslav_solids1,caslav_solids2}. It has also been shown that the thermal environment does not only
tend to destroy QE, but also permanently creates it by mediating between any two systems \cite{Braun1, Braun2}.

Therefore, it is not completely impossible that QE can exist in the biological realm. In fact, its obvious 
advantages may have helped to stabilize mechanisms utilizing QE under evolutionary pressure. 
First suggestions for a role of entanglement between animals have already been made 
in \cite{Josephson}. There have also been hypotheses of quantum computations, which rely on QE,
taking place in the brain \cite{Hagan}. It has also been suggested that correlations found between the 
electroencephalograms from two different persons could be due to QE \cite{Walach, Radin}.
Indeed, one can easily think of a wide range of
biological processes, where QE would lead to a Darwinian advantage: Quantum entanglement could 
coordinate biochemical reactions in different parts of a cell, or in different parts of an organ. 
It could allow correlated firings of distant neurons. And --- as shall be the topic here --- it 
could coordinate the behavior of members of a species, because it requires no physical link and is independent of 
distance. It is also conceivable that QE correlates processes between members of different species, and even 
between living systems and the inanimate world.

Specifically, the evolutionary advantage of quantum entanglement could consist in the following:
Since quantum entanglement involves no transfer of information between the entangled parts, but produces 
effects over arbitrary distances as if information had been transferred \cite{one.bit}, 
it can {\em eliminate the need} to develop further means of communication.

Nevertheless, QE is a fragile relation easily broken in the thermodynamic
environment of living systems. It is therefore reasonable to expect that QE could
have evolved predominantly with such tasks and processes, which can also be accomplished 
{\em without} QE, but which will consume less resources like time or energy, if 
{\em supported} by QE. 

In this paper we want to present two such examples. They deal with the 
cooperation of two ants and of two butterflies, respectively. Their behavior is simulated in a numerical model, 
as is often done in theoretical biology (e.g. \cite{Hemelrijk} and references therein).
Ants and butterflies were chosen, because these small creatures sometimes show impressively coordinated 
achievements which make it hard to believe that solely the limited computational capacity of their nervous 
systems should be responsible for them. We will, however, not specify a physical 
or biological mechanism which carries and protects the systems within an individual which are entangled with 
similar systems in the other individual, although some remarks will be made in the discussion. We 
are here mainly interested in how QE could be embedded in the stream of sensing, 
decision making and acting of the individuals. We shall see that one way how QE could 
work is as an "inner voice", which suggests a specific action to the individual, despite a lack 
of sufficient information for a rational decision. In this sense it could operate like some
quantum games, which help the players win even if there is no communication between them 
\cite{Klimovitch}, so that some of these games exhibit properties of "pseudo telepathy" 
\cite{pseudo.telepathy}.

\vspace{5mm}

{\bf 3. The models}

\vs

{\em 3.1.  Two ants pushing a pebble}

We look at two ants who must push a pebble towards a certain goal. 
Each ant $j$, $(j=1,2)$ is able to push with a force $\vec f_j$. In order to move the pebble a
minimum force $f_{min}$ must be applied. Clearly, if the pebble is too heavy to be moved by any
of the ants alone, i.e. $f_{min} > |\vec f_1|$ and $f_{min} > |\vec f_2|$, 
the two ants must push simultaneously, and they must push in similar directions.

The two ants go about their task by making a series of {\em simultaneous} push attempts. However,
at each push attempt each ant decides with a probability of $\frac{1}{2}$ whether to push or to 
rest. This is where we will permit QE to come in. A push 
attempt is successful if the force applied to the pebble is larger than the required minimum. Then
the pebble will move a little path length proportional to the force and in the direction of the 
force. In order that the two ants shall be able to exploit QE we will speculatively assume that
somewhere in their nervous system they have a region which contains spin-$\frac{1}{2}$ systems, e.g.,
an array of atoms with a magnetic moment. And each of these spins shall be quantum mechanically
entangled with exactly one such spin in the other ant, where any of these pairs shall be in the
triplet state, eq.(4). An ant shall be able to make a quantum mechanical spin
measurement on any of its spins. Somewhat anthropomorphically we can imagine the process of 
measurement to be a moment of introspection whose result the ant notices as a sudden urge 
to do this or that, depending on the outcome of the spin measurement.
 
Each of the pairs of spins shared by the two ants is reserved for a specific push attempt. 
Before a push attempt, an ant must make two decisions:
\begin{itemize}
{
\item Choose the direction $\beta_j$ of the push. This decision is derived from the
sensory input, specifically from the odor emanating from the goal. The chosen direction shall be
subject to a probability distribution $w(\beta_j)$. Thereby we simulate that the ants are not 
clever enough to keep concentrating on the task, or that they may be distracted by other sensory
input, or that gusts of wind may temporarily obscure the olfactory information, etc.
}
{
\item Decide whether to really push at this attempt, or whether to have a little rest.
This decision will be made by a quantum measurement of the spin of the particle
reserved for this attempt along the push direction just chosen. The ant will only push, if
the result is "+". This option simulates an ant's need to come to a decision about an action
despite a general lack of sufficient information, or a general inability to come to an 
informed decision because of the limited capacity of its brain.
}
\end{itemize}
These decisions are made independently by each ant and they are not communicated to the other ant.
Of course, {\em after} the action, each ant could in principle
obtain some information about the decisions of the other ant from the effect on the 
position of the pebble. We assume that this shall not influence an ant's further strategy, because
we wish to see the pure effect of the quantum correlations without 
communication between the partners. (Communication {\em after} the fact  
will be permitted in our next example with the butterflies.)

In case ant $j$ does decide to push, the force it applies to the pebble is
\be
{\vec f}_j = s_j \left( \begin{array}{c}
\sin \beta_j \\
\cos \beta_j
\end{array}
\right),
\ee
where $s_j$ represents the strength of the ant. The direction $\beta_j$ can range from
$-\pi$ to $+\pi$ where $\beta_j = 0$ is the direction straight to the goal.

In a single push attempt there are three possibilities how force can be applied to the pebble. 
Either by one ant alone, which happens with probabilities $p_{+-}^{(t)} = p_{-+}^{(t)} = 
\frac{1}{2} \sin^2\left(\frac{\beta_1 - \beta_2}{2}\right)$ [eq.(6)]. Or by both ants together, which 
happens with probability $p_{++}^{(tt)} = \frac{1}{2} \cos^2\left(\frac{\beta_1 - \beta_2}{2}\right)$ [eq.(5)].
Knowing that in all three cases the pebble will only move if the force is larger than $f_{min}$,
the expected displacement of the pebble after {\em one} push attempt is
\begin{eqnarray}
\vec R & = & g \sum_{j=1,2} {\int_{-\pi}^{\pi} d\beta_j 
w(\beta_j)\frac{1}{2}\sin^2\left(\frac{\beta_j - \beta_{\ne j}}{2}\right)
{\vec f}_j(\beta_j) \Theta \left(\left|{\vec f}_j \right| - f_{min}\right) 
} \nonumber \\ 
 & + &
g \int_{-\pi}^{\pi} d\beta_1 \int_{-\pi}^{\pi} d\beta_2 F(\beta_1,\beta_2),
\end{eqnarray}
where the abbreviation $F(\beta_1,\beta_2)$ stands for
\be
F(\beta_1,\beta_2)= w(\beta_1) w(\beta_2) \frac{1}{2}\cos^2\left(\frac{\beta_1 - \beta_2}{2}\right)
\left[{\vec f}_1(\beta_1) +{\vec f}_2(\beta_2)\right]
\Theta \left(\left|{\vec f}_1 +{\vec f}_2 \right| - f_{min}\right)
\ee
Here, $g$ is the proportionality constant translating applied force to pushed distance in a 
single push attempt, $\beta_{\ne j}$ refers to the angle not integrated over, and $\Theta(x)$ is 
the step function whose value is $1$ if $x\ge 0$ and $0$ otherwise. Since successive push
attempts are independent of each other, the expected endpoint after $N$ push attempts is simply 
$N{\vec R}$.

This model of quantum entangled ants can be contrasted with one of two independent ants, in which the
choice of direction happens exactly as in the quantum model, but the decisions of whether to push or to rest
are made by each ant independently. We can still imagine these
decisions to be based on spin measurements along the chosen directions. But the spins shared by
the two ants will have no QE whatsoever. Then the above quantum mechanical 
probabilities $p_{+-}^{(t)}$, $p_{-+}^{(t)}$, and $p_{++}^{(t)}$ will be independent of the chosen angles, and 
will have a constant value of $\frac{1}{4}$. Note that, in this independent case, the ants will also be
able to push the pebble towards the goal. And we can expect that, as long as the pebble is light 
enough to be pushed by a single ant, there should be little difference in how far the pebble gets 
pushed with a given number of attempts. However, as soon as the effort of both ants is needed to 
move the pebble, the quantum entangled ants will become superior. The reason is that 
especially for small differences between $\beta_1$ and $\beta_2$ -- which are the choices of angles
needed to move a pebble which is too heavy for a single ant -- the ants will make the {\em same} decision of 
whether to push or to rest up to {\em twice as often} as the independent ants. Hence, the behavior 
of the quantum entangled ants will be more coordinated, resulting in fewer futile push attempts. Nevertheless,
the behavior of a {\em single} ant which is quantum entangled with another one, 
will be indistinguishable from a completely independent ant.
\begin{figure}[t]
\begin{center}
\epsffile{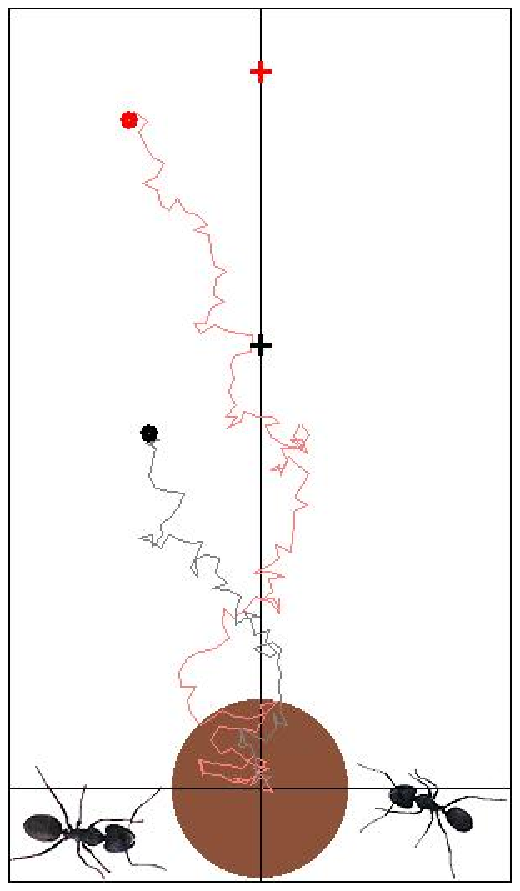}
\caption[Fig.1:]{ Typical stochastic paths of the pebble as pushed by quantum entangled ants (red) as well as
by independent ants (black). In both cases 600 push attempts were made. The crosses show the theoretically expected
endpoints of the pebble. The probability distribution of the push directions was set such that it was three times
as likely for the ants to push forward than to push backward (z=2/3 in eq.(10)). One ant could exert a force of 0.9,
the other a force of 1.1.
}
\end{center}
\end{figure}

Figure 1 shows the result of 600 push attempts of the two ants, for the independent (black) as well as
for the quantum entangled case (red). It is taken from a screen shot of the simulation the actual random
decisions \cite{VBprograms}. The ants must push the pebble to the upper line of the picture. The erratic lines 
with the fat endpoints show typical paths of the pebble. The crosses indicate the theoretically expected 
endpoints of the pebble, as given by eqs.(8) and (9). For the sake of simplicity the probability distribution for 
the choice of push directions of an ant has been assumed to be the same for both ants and was taken
as piecewise linear for positive and negative angles, respectively,
\be
w(\beta) = n\left[\pi-z\left| \beta\right|\right],
\ee
where $n$ is an appropriate normalization factor and $z$ is a positive constant between $0$ and $1$.  
For $z=0$ there is no preferred direction. For $z>0$ the directions towards the goal become more likely and 
for $z=1$ the distribution is triangular with the peak at $\beta=0$ (pointing directly to the goal). 
Figure 2 shows the ratio of the distance the pebble can be pushed by the quantum entangled ants over the
distance achievable by the independent ants with the same number of push attempts. Note that these 
results turn out to be independent of the particular value of $z$ in the probability distribution 
of directions in eq.(10). The solid line shows the case of two equally strong ants (both can exert
a pushing force of 1). Not surprisingly, the superiority of the quantum entangled ants really sets in as
soon as the pebble becomes too heavy to be moved by one ant alone. If the pebble
requires the maximum force of the two ants, the quantum entangled ants can push the pebble twice as
far as the independent ones. The reason is that both ants must push in the same direction. Then the
quantum entangled ants will either both push or both rest, while the independent ants will do so only half 
as often. The more general case of two ants of different strength is shown by the dashed line. 
A certain advantage of the quantum entangled ants sets in as soon as the force needed
to move the pebble is larger than can be exerted by the weak ant (which here is 0.9), because 
all efforts by the weak ant alone become futile. If the pebble becomes so heavy that even the 
strong ant cannot move it alone, the full advantage of the quantum entangled ants sets in. 

It is interesting to note that in both scenarios a very small advantage for the entangled ants 
exists also when the pebble can be moved by any of the ants alone. This is due to the fact that, 
even for light pebbles, there is a difference in how the pebble gets moved. E.g., for large 
angles between the chosen directions the forces on the pebble may almost cancel each other 
and the pebble will not move. Such futile attempts occur with some frequency for
the independent ants. But they are rare for the quantum entangled ants, because the 
$\cos^2(..)$-term in eq.(3) tends to vanish for such angles. Thus quantum entangled ants profit from
making fewer "wrong decisions".
\begin{figure}[ht]
\begin{center}
\epsffile{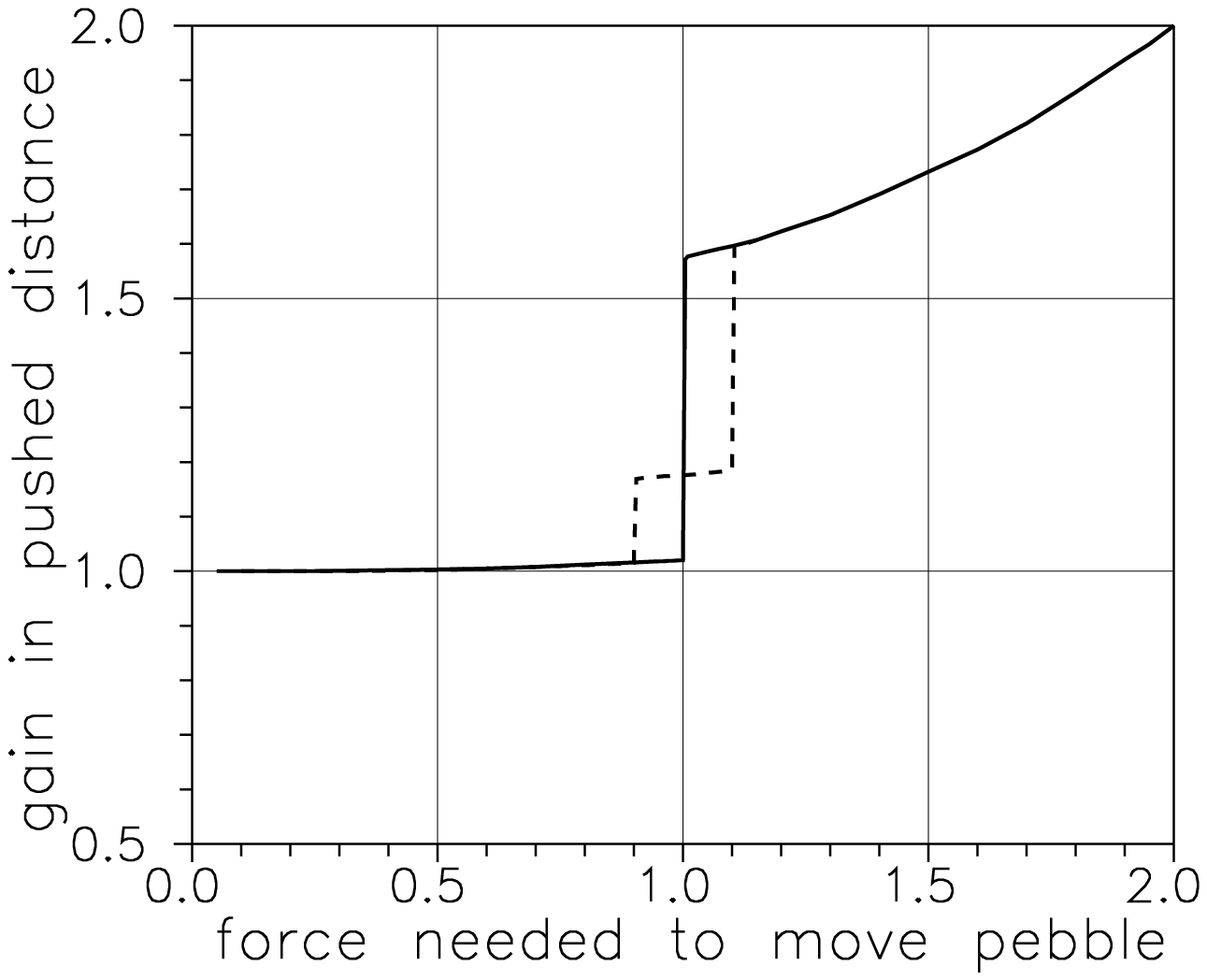}
\caption[Fig.2:]{ 
Gain in distance the pebble can be pushed by the quantum entangled ants relative to the
independent ants, as a function of the force needed to move the pebble. Solid line: Both ants 
have force 1. Dashed line: One ant has force 0.9, the other has force 1.1.
}
\end{center}
\end{figure}

\vs

{\em 3.2.  Two butterflies finding each other}

With certain kinds of butterflies it is known that a male and a female can find each other even 
when they are initially many kilometers apart. The usual explanation is that each butterfly
emanates scent molecules to guide the other one. The huge antennas of a butterfly capture the 
molecules, permitting it to determine the gradient of the distribution and hence the
direction of the origin of the scent. Nevertheless, one may wonder, whether for large distances 
the information contained in the few scent molecules is sufficient to give a butterfly 
a clear direction where to fly.

In the present scenario we shall assume that this information will allow a butterfly 
only  to come up with a probability distribution for the directions in which it should fly. The decision
whether the butterfly will actually do a short flight in the chosen direction will come from a
quantum measurement. As in the case of the ants, we assume that the two butterflies share a large number of  
maximally entangled pairs of spin-$\frac{1}{2}$ particles, of which each butterfly holds one 
particle. However, here the entangled state will be the singlet state (eq.1),
because we want the butterflies to get the same measurement result, if they measure their respective spins 
along opposite directions.

The details of the scenario are as follows:

\noindent
(i) The intensity of the scent emanated by each butterfly drops off as $1/r^2$ where $r$ is the 
distance from the butterfly.

\noindent
(ii) The propagation of the scent is very much faster than the speed with which the butterflies fly, 
so that each butterfly can notice a change of the distance of the other one with little delay 
as a change of intensity of the scent.

\noindent
(iii) Each butterfly moves in a sequence of short straight flights of constant length. Before such a 
short flight the butterfly has to make two decisions in the following order:
\begin{itemize}
{\item Choose a direction for the short flight.}
{\item Decide whether to really do the short flight, or whether to have a little rest.}
\end{itemize}
The first decision is resolved in the usual neuronal manner: The butterfly chooses the direction 
for the short flight randomly, but weighted with the probability distribution of directions which 
it considers appropriate in view of its experience of change of intensity of the scent in the previous short 
flights. In the model calculations, each butterfly can choose among 16 directions evenly spaced 
over $2\pi$. In the beginning, this probability distribution is isotropic. After each short flight,
the distribution is updated according to a rule which will be explained later.

\noindent
The second decision comes from a quantum measurement. The butterfly measures the 
spin-$\frac{1}{2}$ particle designated for this short flight along the chosen direction. If the 
result is "+", it does the short flight, otherwise it rests until the next short flight is due.

\noindent
The rule for updating the probability distribution of flight directions now looks as follows. 
(It is only applied, if the short flight has actually taken place. If, instead, the butterfly has 
taken a rest, it will retain the probability distribution from before the rest.) The butterfly 
measures the intensity of the scent of the other one:
\begin{itemize}
{\item 
If the increase of the intensity, i.e. the average gradient of the scent along the short flight 
path, is above a certain threshold, the butterfly judges this to have been a good direction and 
enhances the corresponding probability weight by the factor $(1+\l)$. This direction is then more 
likely to be chosen again in one of the next short flights. The parameter $\l$ can be set between 0 
and 1. When it is 0, no learning from experience occurs.
}
{\item
If the increase of the intensity is below the threshold, the butterfly flies back, because 
it judges this to have been a bad direction. In addition, it reduces the probability 
weight of this direction by the factor $(1+\l)^{-1}$. This direction is then less likely to be 
chosen again in one of the next short flights.  
}
\end{itemize}
The threshold is taken as a certain fraction of the strongest increase of the intensity of the 
scent encountered in the short flights until then. Therefore, as the butterflies get closer to each 
other, the threshold will get higher and they will become more discriminating in judging a short 
flight as having been good or bad.

This quantum scenario can again be compared to a scenario of butterflies who make independent decisions.  
In that scenario the decision before each short flight, whether to fly or to rest, is made 
completely independently by the two butterflies. Each will decide randomly with a constant 
probability of 0.5 whether to fly or to rest.
\begin{figure}[ht]
\begin{center}
\epsffile{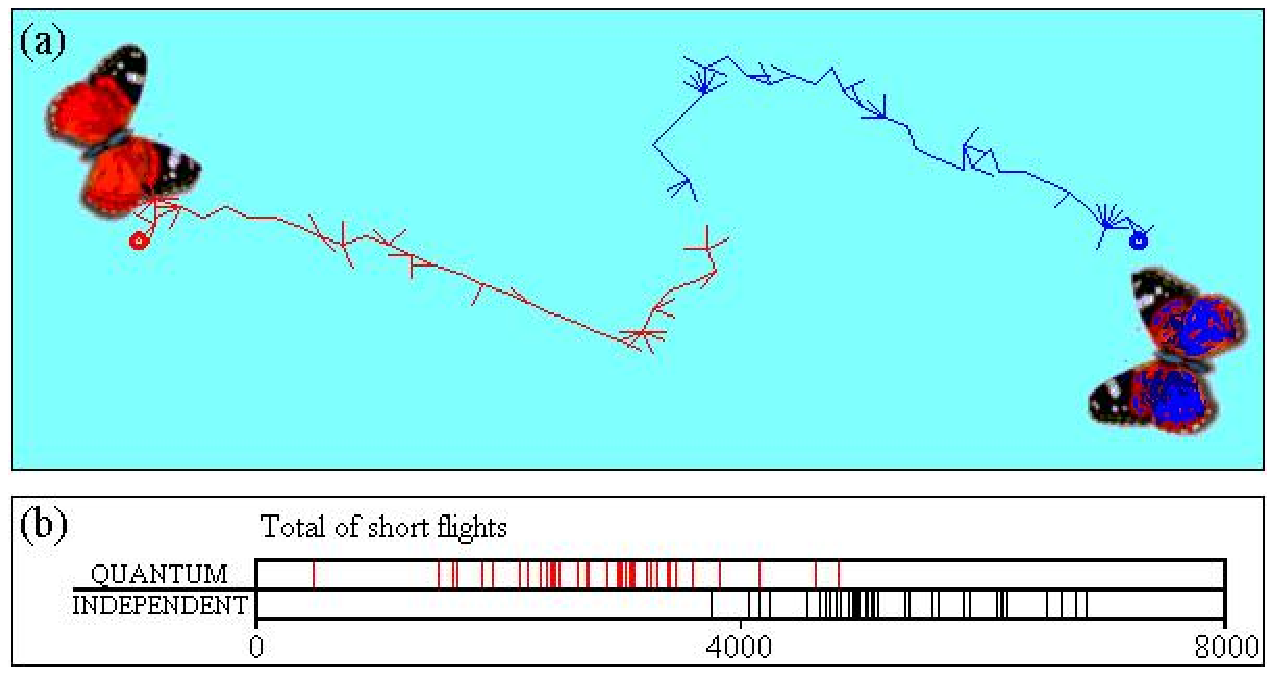}
\caption[Fig.3:]{ 
{\bf (a):} Typical flight paths of the two butterflies. Initial distance: 1600 units. Length of one 
short flight: 40 units. Learning factor $\l=0.5$.
The lines jutting out from the two main lines are short flights after which the respective 
butterfly had to fly back. Note that sometimes several attempts had to be made to find a good
direction.
{\bf (b):} Statistical results of 40 independent and 40 quantum runs, respectively. 
Initial distance: 1600 units. Length of one short flight: 5 units. Learning factor $\l=0.5$. 
The quantum entangled butterflies needed an average of 2778 short flights to find each other. 
The independent butterflies needed an average of 5255 short flights.
}
\end{center}
\end{figure}

Figure 3 shows parts of the screenshot of the program which simulates the behavior of the
two butterflies \cite{VBprograms}. The top picture (a) demonstrates actual tracks. The 
interesting result is the total number of short flights including the back flights, i.e., the total
flight distance the two butterflies had to cover to meet. This is shown in the lower picture (b)
which indicates the statistical results of 40 runs of independent, and of 40 runs of quantum
entangled butterflies, respectively. The red pointers in the upper 
half indicate the number of short flights of quantum entangled butterflies, the black pointers in 
the lower half those of the independent butterflies. It is noticeable that the 
quantum entangled butterflies can find each other with much fewer short flights than the independent ones. 
This is because the former decide more often simultaneously to actually do the short flight, if 
the chosen directions happen to point roughly towards one another. This in turn, gives more often 
valid short flights, i.e. short flights where the increase of the scent is above the threshold so 
that neither of them will have to fly back. Also, it leads to a quicker adaptation of the 
probability distribution of flight directions to favour the current good directions.

The derivation of the theoretical expression for the number of short flights needed until the encounter 
takes place is a laborous exercise and will not be given here. Instead, Figure 4 shows for 40 quantum and 
for 40 independent runs, respectively, the averages and standard deviations of the total flights needed 
until the butterflies meet, as a function of the learning factor $\l$. The initial distance of the 
butterflies is 1600 units. The length of one short flight is 5 units. 
The threshold for flying back is that the butterfly does not undo a short flight if the increase of the 
intensity of the scent of the other butterfly was at least 60\% of the strongest increase of the scent 
found until then. Obviously, the quantum entangled butterflies find each other faster. For a learning 
factor of $\l=0$ they need only about 83\% of the short flights of the independent butterflies, and for 
$\l=1$ they need only about 48\%.
\begin{figure}[ht]
\begin{center}
\epsffile{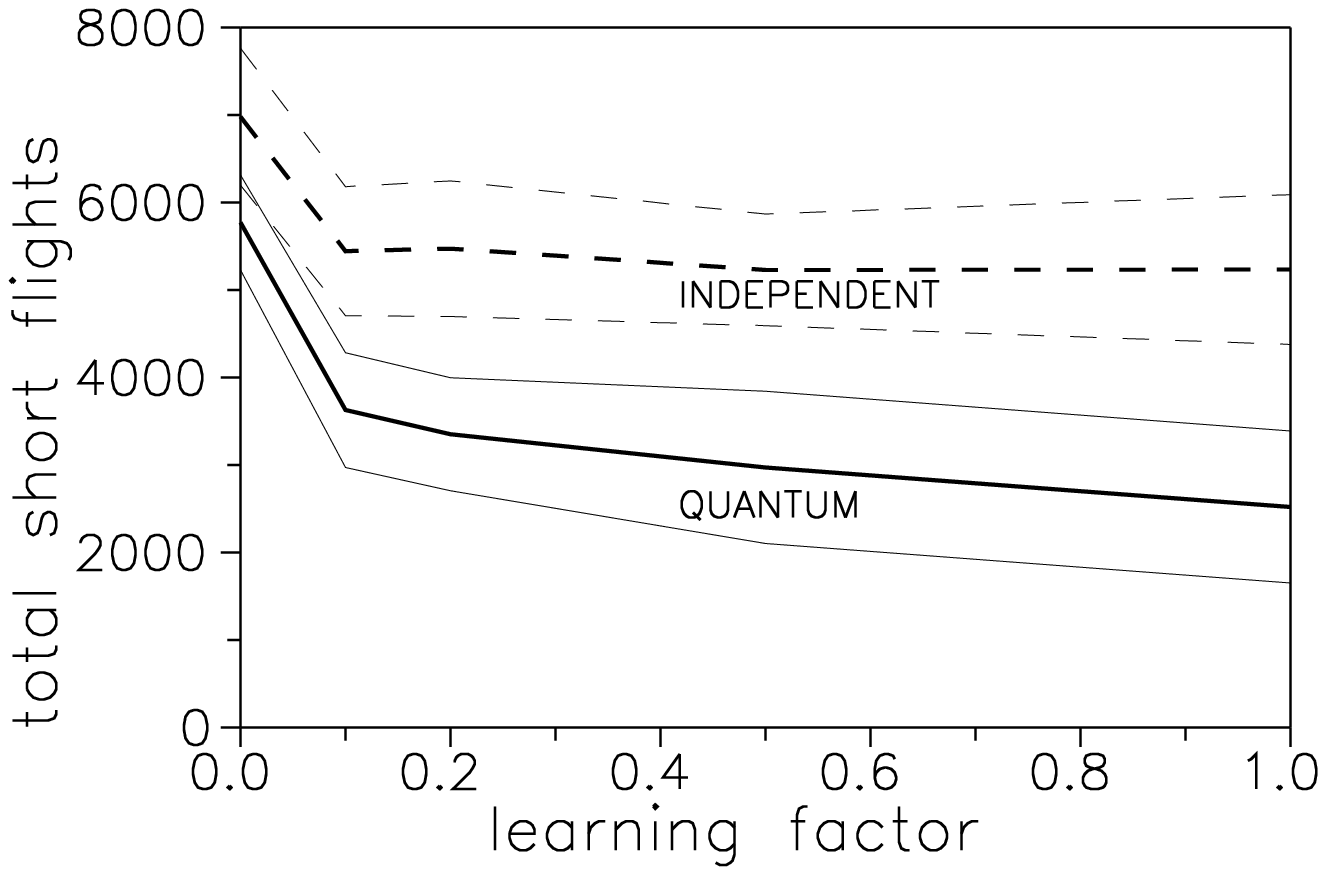}
\caption[Fig.4:]{ 
Total number of short flights needed by the two butterflies to find each other, as a function
of learning factor $\l$. Calculated from 40 runs at each $\l$. Solid lines: Average and range of 
standard deviation for quantum entangled butterflies. Dashed lines: For independent butterflies. 
}
\end{center}
\end{figure}

It may be surprising that the quantum entangled butterflies have an advantage 
over the independent butterflies even for $\l=0$, i.e., when the butterflies don't learn from
experience so that for all short flights all directions remain equally likely. 
This can be understood qualitatively when looking at the probabilities that a 
butterfly does {\em not} have to fly back after a short flight. Suppose that for 
the next short flight the butterflies happen to choose directions pointing exactly at each other. 
If both happen to really do the short flight, neither will have to fly back, 
because the increase of the intensity of the mutual scents will be the highest possible. Now, 
the probability that both will fly is $\frac{1}{2}$ for the quantum case, but only $\frac{1}{4}$ 
for the independent case. For the independent case, there is also
a chance of $\frac{1}{2}$ that only one butterfly will fly, but in these instances the increase of the
intensity of the scent will likely not be high enough and the butterfly will fly back. Altogether,
there should thus exist an advantage for the quantum entangled butterflies. And they should have a
similar advantage over quite a range of roughly forward pointing directions, the exact angular 
width of this range depending on the choice of threshold for flying back.

\vspace{5mm}

{\bf 4. Discussion}

Several aspects of the two examples deserve further discussion.

\vs

{\em 4.1. Communication} 

In the example with the ants no communication between the ants is necessary.
At each attempt each ant just selects a direction and does or does not push. The momentary position
of the pebble, from which some information about the previous decisions of the other ant could be
gleaned, has no influence on what choices the ant makes. However, an important communication, which
in practice will be olfactory, must permanently happen from the goal to the ants. A scent must always tell
the ants where the goal is, because the advantage of the quantum entangled ants pushing a 
heavy pebble results solely from their preference of pushing towards the goal: 
For similar choices of directions it leads to more frequent occurrences of both ants' 
{\em simultaneous} decisions to push, which is the only way to move a heavy pebble. 

In the example with the butterflies there is no external goal where both must fly, but each 
butterfly represents the goal for the other one. Therefore, the communication must happen 
between them, as is effected by each one permanently checking the intensity of the scent of the other 
one. Without this information the butterflies would have no preference for flying towards each other,
because the quantum correlations would also favor directions which lead directly away from each 
other. 

\vs

{\em 4.2. Strength of correlations} 

In both examples we have only looked at maximally
quantum entangled decisions of the two cooperating insects on the one hand, and compared them to 
completely independent decisions on the other. Clearly, there is also the possibility of correlations whose 
strength lies anywhere between, e.g., correlations covered by classical physics which would fulfill Bell's 
inequality, and which could be realized by neuronal circuits exploiting all the information held by an individual. 
Such correlations would lead to achievements between those of the two extremes we have investigated,
because the strength of the correlations is directly reflected by how far a heavy pebble gets 
pushed by the two ants, or by how soon the two butterflies find each other, respectively.

\vs

{\em 4.3. Critique of the models} 

The assumptions in our simulations are certainly much too simple
and much too technical to be found in real insects. It is unlikely that living systems store 
spin-$\frac{1}{2}$ particles and that each of these spins is quantum entangled with a similar spin in
another living system. It is even more unlikely that in successive decisions an individual resorts
to just that spin which is entangled with the spin which the other individual happens to resort
to for just the appropriate decision. Such clock-like synchronicity could perhaps be implemented in
artificial devices, but would not appear in animals. Also, there is no reason why entanglement
should exist between just two, instead of three or more individuals. Nevertheless, our simulations 
underline that QE between decisions of cooperating individuals --- no
matter how these correlations are physically realized --- can enhance the cooperative achievements 
significantly beyond those obtainable with even the fullest exploitation of the information available 
to an individual as might be facilitated by neural networks. This is because QE 
does contain extra information which is inaccessible to any individual alone but comes 
to the fore in the result of joint measurements on the quantum entangled systems. 

\vs

{\em 4.4. Creation and persistence of QE}

Having made the above criticisms we mention a few results, which 
might lend support to the existence of QE within or between living systems, because they adress the two
important issues of creation and persistence of the entanglement. The references are only an 
entry to the literature.

\noindent
Let us first look at the creation of QE. QE arises always when two systems interact directly and when
the result of the interaction permits several different outcomes \cite{Zhang}. This mode is more or less 
excluded in our models, because we wanted to treat QE as accessed through a kind of organ and thus 
assumed that the two mutually entangled parts in the insects are protected.
The relevant mechanism must therefore work through intermediaries. This has been much studied in 
the context of quantum computation. One such possiblity is entanglement swapping 
\cite{entanglement_swapping}. It requires auxiliary systems which are already entangled, and which
then interact with the systems to be entangled. The process can be repeated arbitrarily, but the
final entanglement gets more and more diluted. Another possibility is that the systems to be entangled
each interact in succession with one external system. This has been done with two atoms passing a
cavity one after the other \cite{Phoenix}, and with two macroscopic Cs-samples, which one after the 
other were traversed by a beam of light \cite{Julsgaard}. A related possibility is that the
third system interacts independently with the two systems to be entangled, and is then subjected to 
frequent measurements later on \cite{L.-A.Wu}. Here one can also have more intermediate systems. This 
is a favourable mechanism, because the systems which lie between the two entangled parts in the 
two insects will be measured permanently by the thermal environment. In fact, it is now known that the
thermal bath, which surrounds any physical system and even more so any living system, 
also acts as a permanent creator of QE between physical systems \cite{Braun1, Braun2}, and not only as a 
destructive influence as has mostly been assumed.

\noindent
Now to the issue of persistence, which means the slow-down of decoherence. That decoherence, and thus the 
destruction of QE, need not set in immediately has been estimated in a different context for the 
microtubuli of the brain, where quantum states can exist for relatively long periods 
\cite{Hagan,Penrose}. Recently, it has been found for certain kinds of entangled states that the
entanglement of M subsystems, each consisting of many spin-$\frac{1}{2}$ particles, becomes more 
robust against destructive interactions with the environment, when the number of particles per 
subsystem increases \cite{Bandyo}. The most robust entanglement is obtained when
there are only M=2 subsystems. (The experiment showing entanglement between two macroscopic samples
of Cs-atoms seems to confirm this \cite{Julsgaard}). For our models this could mean that 
entanglement between just two living systems is more likely than between three or more systems, 
and that the entangled systems should have many degrees of freedom, which is certainly the case 
for bio-molecules. The findings of Zukowski et al \cite{Marek} and of Mermin \cite{Mermin} point 
in the same direction. Both works show that the violation of Bell inequalities, and thus the 
strength of QE, becomes more pronounced with higher dimensional systems.
A further result, which shows that quantum states need not  
decohere immediately in a thermal environment, comes from the study of the state of a central
spin coupled to a spin bath \cite{Tessieri,Dawson}: If there is strong entanglement 
{\em within} the spin bath, the initially pure state of the central spin will decohere into a 
mixed state only slowly. 
 
\noindent
Even if the questions of creation and protection of entanglement are solved, there is still the
problem that in our models sequential and synchronized measurements of the entangled spins were
needed.
For entanglement to work in living systems, one would probably need very massive entanglement,
such that it is possible for one animal to blindly probe any tiny fraction of its part of the 
entangled system and for the other animal to do the same on any other tiny fraction, and yet a
strong correlation of the results should obtain. It is conceivable that, due to conservation of 
angular momentum, specific configurations of many spins exhibit such properties. 
But details remain to be worked out.

\vspace{5mm}

{\bf 5. Conclusion}

We have modeled the chain of {\em sensing, deciding and acting} in the cooperation of two insects 
of a given species in two different scenarios. We have seen that the quantum mechanical 
correlation of the statistical decisions of the two individuals is clearly an advantage, because 
it eliminates the need for more complex communication between them. Nevertheless, it must be 
realized that actual processes
in nature will involve many more parameters than our models. The value of the models is rather 
instructive: The models illustrate that, in moments of hesitation and indecision due to a lack of 
information, an "inner voice" in the form of the result of a quantum measurement can be of great 
help to an individual in achieving a goal of common interest to the species. Therefore, there may 
have been evolutionary pressure to rely on quantum entanglement for certain decisions individuals 
need to make in the absence of sufficient information.

\vspace{5mm}

{\bf Acknowledgment} 

I am pleased to thank C. Brukner, G. Krenn, K. Svozil and V. Vedral for 
fruitful comments and input.

\end{document}